\documentclass[useAMS,usenatbib]{mn2e}

\usepackage{graphicx}
\usepackage{natbib}

\title[Mid-infrared PL relations for  Globular Cluster RR Lyrae]
{Mid-infrared PL relations for  Globular Cluster RR Lyrae}
\author[A. K. Dambis, A. S. Rastorguev and M. V. Zabolotskikh]{A. K. Dambis\thanks{E-mail:
mirage@sai.msu.ru}, A. S. Rastorguev and M. V. Zabolotskikh\\
Sternberg Astronomical Institute, Lomonosov Moscow State University, Universitetskii pr. 13, Moscow, 119992 Russia}
\begin{document}
 
\date{Accepted 2014 January 31.  Received 2014 January 21; in original form 2013 October 14}

\pagerange{\pageref{firstpage}--\pageref{lastpage}} \pubyear{2013}

\maketitle

\label{firstpage}

\begin{abstract}
The period - metallicity - WISE W1- and W2-band luminosity 
relations are derived for RR Lyrae stars based on WISE epoch photometry for
360 and 275 stars in 15 and 9 Galactic globular clusters, respectively. Our 
final relations have the form $<M_{W1}>$=$\gamma_{W1}$ -(2.381$\pm$0.097)log $P_F$ + 
(0.096$\pm$0.021)[Fe/H] and $<M_{W2}>$=$\gamma_{W2}$ -(2.269$\pm$0.127)log $P_F$ + 
(0.108$\pm$0.021)[Fe/H], where [Fe/H] values are on the scale of
\citet{carretta}. We obtained two appreciably 
discrepant  estimates for the zero points $\gamma_{W1}$ and $\gamma_{W2}$ of  
both relations: one based on a statistical-parallax analysis -- $\gamma_{W1}$=-0.829$\pm$0.093 
and $\gamma_{W2}$=-0.776$\pm$0.093 and 
another, significantly brighter one, based 
on HST FGS trigonometric parallaxes -- $\gamma_{W1, HST}$=-1.150$\pm$0.077 
and $\gamma_{W2, HST}$=-1.105$\pm$0.077. The period-metallicity-luminosity relations in the two bands
yield highly consistent distance moduli for the calibrator clusters and the
distance moduli computed using the W1- and W2-band relations with the HST zero points agree well
with those computed by \citet{sollima} based on their derived
period-metallicity-K-band luminosity relation whose zero point is tied
to the HST trigonometric parallax of RR Lyrae itself 
($\Delta DM_0$ = 0.04 and 0.06, respectively, with a scatter of only 0.06).
\end{abstract}


\section{Introduction}

RR Lyrae variables are known to obey rather tight period-metallicity-luminosity relations of the form
\begin{equation}
<M_X> = \alpha_X \cdot log P_F+\beta_X \cdot [Fe/H] + \gamma_X,
\label{PMLR}
\end{equation}
where $<M_X>$ is the intensity-mean absolute magnitude in the photometric band X and
$P_{F}$ is the fundamental-mode period (equal to the variability period $P$ for RRab type variables, which pulsate in
the fundamental mode and log $P_F$ = log $P$ +0.127 or $P_F = P$/0.746 -- for RRc type variables, which pulsate in the first overtone)
in various photometric bands~X \citep{catelan}. (The above first-overtone to fundamental period ratio 
dates back to theoretical estimates by \citet{iben} and is commonly used by most of the authors to fundamentalise the periods of RRc type
variables -- see, e.g., \citet{fs, sollima, flkvw}. Earlier model-based estimates  yield a
period conversion factor corresponding to log $P_F$ = log $P$ +0.130 \citep{VAB}. More recent stellar models corroborate these results
and, as \citet{castellani} point out, show that the adopted procedure yields fundamentalised 
periods with an uncertainty no larger than $\delta log P_F$~=~$\pm$~0.005 \citep{bono,  marconi}. Further support
for the small uncertainty is provided by the observed period ratios of double-mode
RR Lyrae type stars (RRd) - as is evident from the Petersen diagram for Galactic and LMC RRd type stars shown in Fig.~2
in \citet{poleski}, the period ratios in all these objects are constrained to the narrow interval from 0.742 to 0.748 corresponding
to the interval of logarithmic corrections from +0.126 to +0.130). It is these relations that make RR Lyraes very popular standard candles used extensively to estimate
distances to stellar systems harbouring old populations. Recently, mid-infrared light curves have been
acquired for several thousand RR Lyraes as a result of spaceborne WISE all-sky photometric survey \citep{wise},
and hence establishing the period-metallicity-luminosity relations for  these stars at least in some  of the WISE bands 
has become a task of prime importance. The progress so far achieved in this direction includes
(1) a study by \citet{klein}, who found $\alpha_{W1} =  – (1.681 \pm 0.147)$ 
with no evidence for metallicity term $\beta_{W1}$ by computing  posterior distances of 76 well observed RR Lyrae based on
the optically constructed prior distances; (2) a conclusion by \citet{dambis3} that the period and metallicity slopes
of the $W1$-band PML relation are practically identical to those of the $K_s$-band PML relation ($\alpha_K$=$\alpha_{W1}$=-2.33
and $\beta_K$=$\beta_{W1}$=+0.088) based on the small scatter of the estimated $<K_s>-<W1>$ 
intrinsic colour indices of Galactic field RR Lyraes with known metallicities, and (3) the study of \citet{madore},
who derived WISE $W1$, $W2$, and $W3$-band RR Lyrae PL relations  
based on the trigonometric parallaxes of four Galactic field RR Lyraes. The problem with the results of \citet{klein} is that
these authors do not fundamentalise the periods of c-type RR Lyraes, which is evidently a bad idea given that RRc type stars 
form a well-defined $\Delta$~log($P$)=-0.127 period-shifted branch of the PL relation in the $K$ band ($\lambda_{eff} \sim 2.2 \mu m$)
and there are no reasons for RR Lyrae variables to behave differently in the W1 band ($\lambda_{eff} \sim 3.4 \mu m$).
The conclusion of \citet{dambis3} might not be entirely correct, because the small scatter of computed $(<K_s>-<W1>)_0$)
intrinsic colour indices may be a result of the star-to-star variations of the period and metallicity terms cancelling
each other because of the appreciable correlation between $log P_F$ and [Fe/H]. Finally, the slopes of the 
$W1$, $W2$, and $W3$-band PL relations estimated by \citet{madore} have large errors because of the very small number 
of stars involved (four). It therefore makes sense to try to estimate the period slopes of the RR Lyrae PML relation  
in some of the WISE bands in a way that would eliminate the effect of metallicity term. 

In this study we follow the footsteps of \citet{sollima} and use photometric data for RR Lyrae variables in 
globular clusters to derive the period slopes ($\alpha$) for the RR Lyrae PML relation in the WISE W1 and W2 bands, because,
as the above authors point out, "The advantage of using GCs in constraining the coefficients $\alpha, \beta$ and
$\gamma$ lies in the fact that all the stars in a given cluster are 
at the same distance, and can be considered to share the same metal content and be 
subject to the same extinction effect."
We then estimate the corresponding metallicity slopes ($\beta$) of these relations based on photometric data for field RR Lyrae type 
variables with known [Fe/H], and, finally, infer the zero points ($\gamma$) of the corresponding relations based 
on (1) results of statistical-parallax-analysis by \citet{dambis3} and (2) HST FSG trigonometric parallaxes.

\section{The data}

\begin{figure*}
 \includegraphics[width=17.cm]{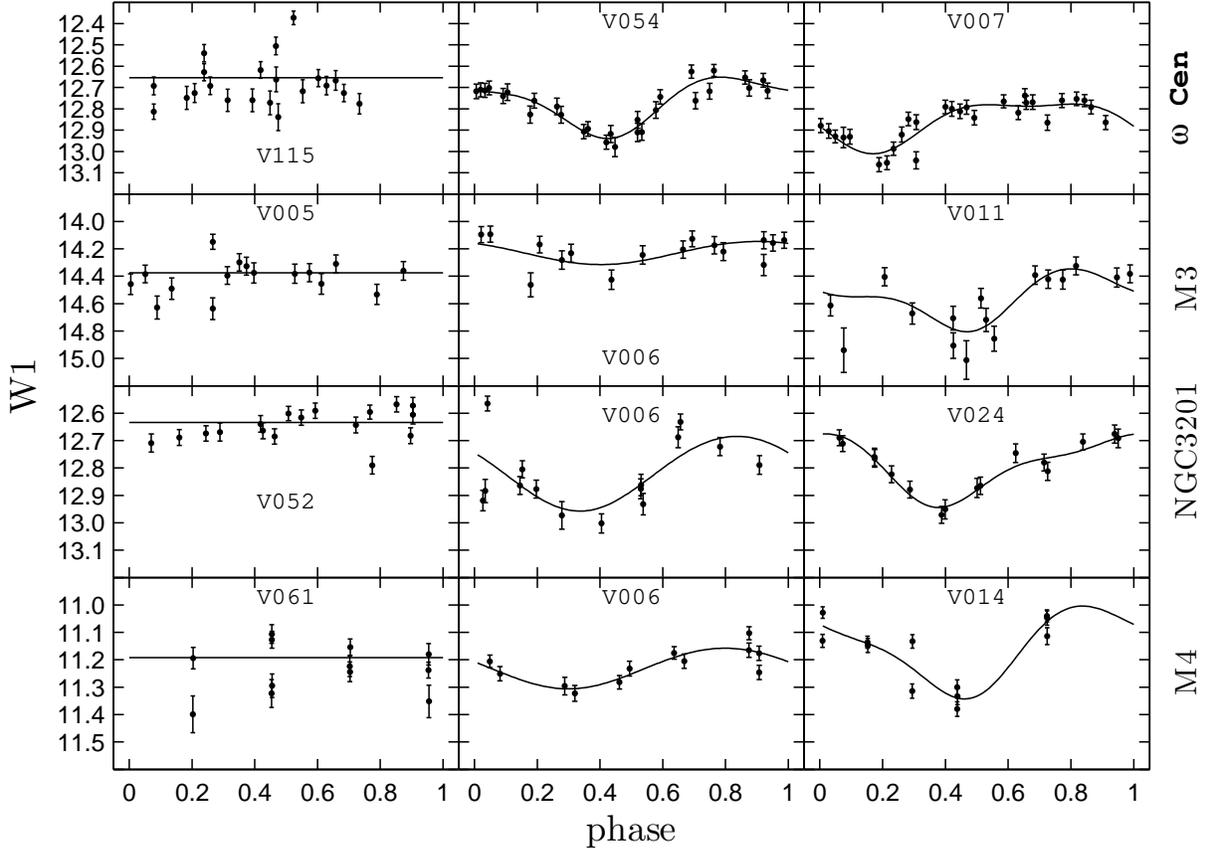}
\caption{Examples of  RR Lyrae type star W1-band light curves of various quality in four globular clusters.}
\label{w1}
\end{figure*}

\begin{figure*}
 \includegraphics[width=17.cm]{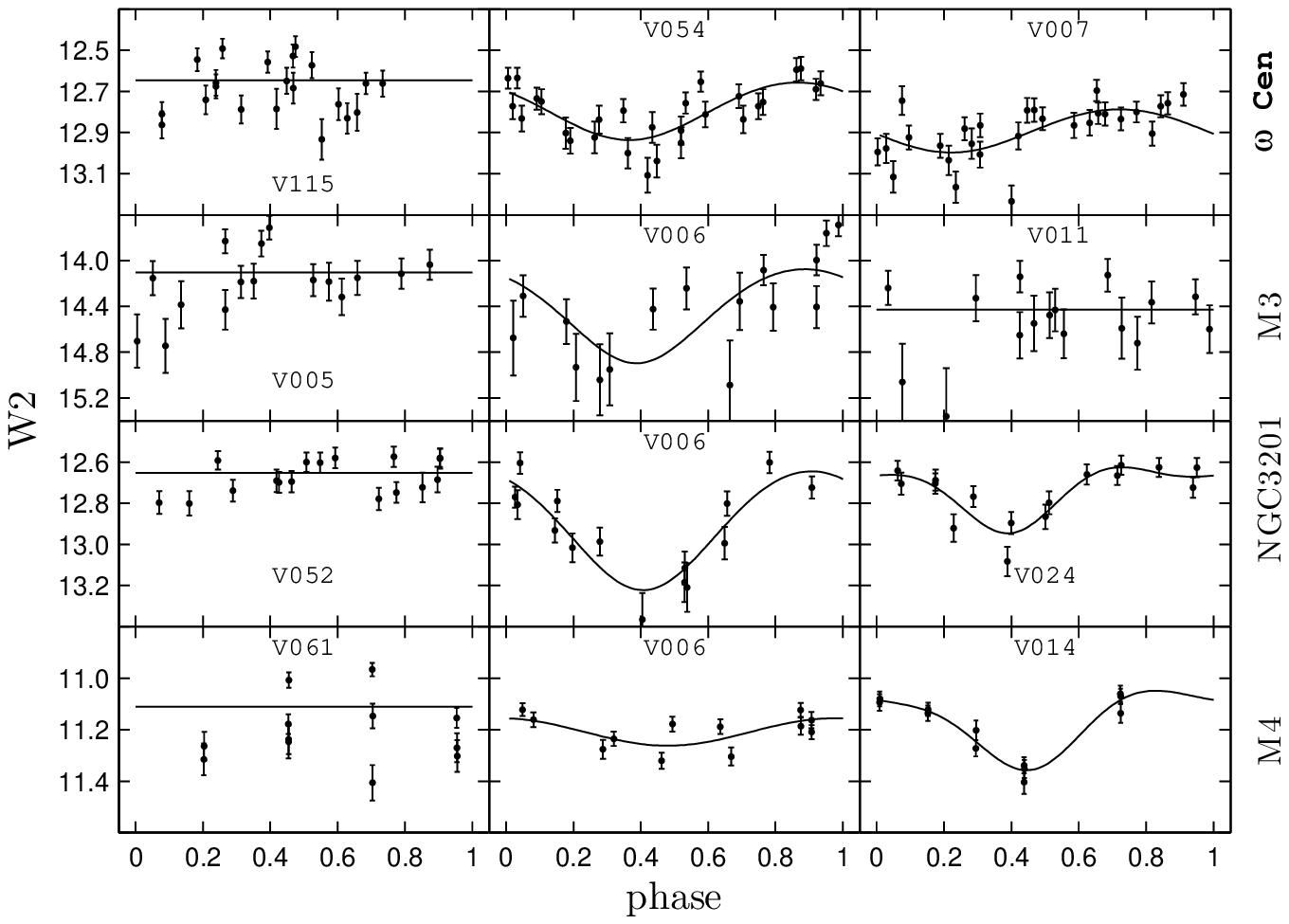}
\caption{Examples of  RR Lyrae type star W2-band light curves of various quality in four globular clusters.}
\label{w2}
\end{figure*}

Last year, the WISE All-Sky Data Release \citep{cutri_wise} was made public, mapping 
the entire sky in four mid-infrared bands W1, W2, W3, and W4 with the effective wavelengths 
of 3.368, 4.618, 12.082 and 22.194~$\mu$m, respectively \citep{wise}. We cross-correlated 
the WISE single-exposure database with the Catalogue of Galactic globular-cluster variables
by \citet{clement}, the Catalogue of Accurate Equatorial Coordinates for Variable Stars in Globular Clusters
by \citet{samus}, and the catalogue of \citet{SH} (for $\omega$~Cen, NGC6723, and NGC6934) 
to  compute (via Fourier fits) the intensity-mean average W1- and W2- band 
magnitudes, $<W1>$ and $<W2>$, for a total of 357 and 272 RR Lyrae type variables in 
15 and 9 Galactic globular clusters, respectively. 
Figures~\ref{w1} and \ref{w2} show examples of W1- and W2-band light curves of different
quality. As is evident from these samples, the phase coverage is more or less satisfactory 
in most of the cases, although the quality of the light curves differs greatly. The
order of the Fourier fit naturally depends on the light-curve quality with only the constant term is left for the
poorest curves. 

The  list of 360 globular-cluster RR Lyrae type stars used in this study is presented in
Table~\ref{mastertable} (its full  version  will be available from the CDS). The columns of this table provide
the following information: (1) NGC designation of the cluster; (2) other commonly used name of the cluster;
(3)  name of the variable; (4)  variability period in days;
(5) W1-band  intensity-mean magnitude with (6) its standard error; (7) W2-band intensity-mean  magnitude with 
(8) its standard error; (9) variability type (RR0, RR1, and RR2 indicate type ab, c, and d variables, respectively, and
RR9 indicates variables with unknown subtypes),
and (10) a flag indicating whether the particular variable was used in the final PL relation
fit (1 - used and 0 - rejected). 

\begin{table*}
 \centering
  \caption{The data for RR Lyraes in the calibrator GCs. This is a sample of the full version, 
which is available in the online version of the article (see Supporting Information).}
  \begin{tabular}{@{}llllcccccc@{}}
  \hline
  Cluster  & Alternative  & Variable    & Period,  & $<W1>$ & $\sigma <W1>$ & $<W2>$ & $\sigma <W2>$ & Type & Use/ \\
  name     & cluster name & name        & days     &        &               &        &               &      & Reject \\
 \hline
NGC3201 &           &  V003 &  0.5994    &   12.3083 &  0.0201  &  12.3628 &  0.0239  &   RR0  &   0 \\
NGC3201 &           &  V004 &  0.6300    &   12.6777 &  0.0145  &  12.6880 &  0.0235  &   RR0  &   1 \\
NGC3201 &           &  V006 &  0.5253    &   12.8124 &  0.0286  &  12.8956 &  0.0269  &   RR0  &   1 \\
NGC3201 &           &  V007 &  0.6303    &   12.6029 &  0.0090  &  12.6673 &  0.0230  &   RR0  &   1 \\
NGC3201 &           &  V008 &  0.6287    &   12.5628 &  0.0333  &  12.5852 &  0.0581  &   RR0  &   1 \\
NGC3201 &           &  V009 &  0.5255    &   12.6367 &  0.0150  &  12.6491 &  0.0253  &   RR0  &   1 \\
NGC3201 &           &  V010 &  0.5352    &   12.7077 &  0.0203  &  12.6578 &  0.0242  &   RR0  &   1 \\
NGC3201 &           &  V011 &  0.2990    &   11.4651 &  0.0118  &  11.5245 &  0.0304  &   RR1  &   0 \\
NGC3201 &           &  V012 &  0.4956    &   12.7398 &  0.0199  &  12.8645 &  0.0361  &   RR0  &   1 \\
NGC3201 &           &  V013 &  0.5752    &   12.4619 &  0.0219  &  12.4421 &  0.0234  &   RR0  &   0 \\
\hline
\end{tabular}
\label{mastertable}
\end{table*}

\begin{figure*}
 \includegraphics[width=17.cm]{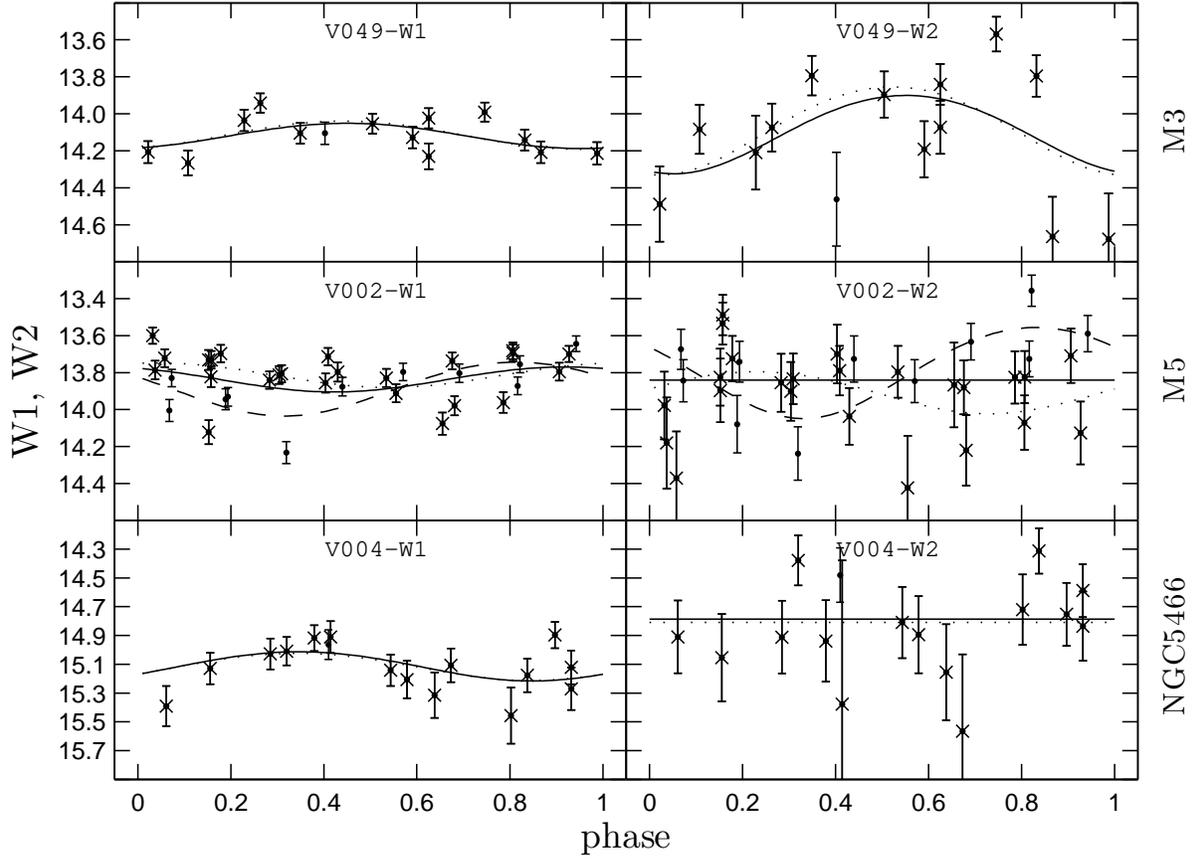}
\caption{Examples of  W1- (left) and  W2-band (right) light curves of some Blazhko RR Lyrae variables in M3, M5, and
NGC5466. The crosses and dots show the measurements corresponding to the "first" and "second" epoch intervals, respectively.
The dotted and dashed curves show the light-curve fits based on the "first" and "second" epoch intervals, respectively, and
the solid curves, the light-curve fits based on all available measurements.}
\label{blazhko}
\end{figure*}

A potential source of error is the Blazhko effect -- long-period variations of the form and amplitude of the light curve --
exhibited by some RR Lyraes. There are known Blazhko stars
in five clusters of our list: M3, M5, M15, NGC3201, and NGC5466. The Blazhko effect should not introduce appreciable 
errors in the computed intensity-mean magnitudes for stars in M15 and NGC3201 because  the time span covered by
WISE observations in these clusters ($\sim$~1.1 and 3.9~days, respectively) is short compared to typical Blazhko periods,
which are on the order of several dozen days. Each of our RR Lyr star in M3 has 14 WISE measurements including 12 
observations concentrated within a $\sim$~1.4-day interval (MJD 55375.074090--55376.463216) and two 
observations near MJD 55203.412309. However, we found that the inclusion/exclusion of the two "outlying"
observations has no appreciable effect on the computed intensity-mean averages in either W1 or W2
with the differences not exceeding 0.009$^m$ and 0.046$^m$, respectively (the standard errors of
the computed intensity means are greater than 0.012$^m$ and 0.042$^m$ in W1 and W2, respectively,
for all the stars concerned). WISE observations of RR Lyraes in M5 were made within two epoch intervals 
(MJD 55411.716751 -- 55412.708967 and 55231.073846 -- 55234.315989) including 12 and 22 measurements,
respectively. The light curves for the two intervals differ appreciably, and
the computed intensity means differ by less than 0.067$^m$ and 0.130$^m$ in W1 and W2, respectively.
The intensity means based on all observations and computed ignoring the variation of the light-curve  shape
and amplitude differ from the intensity-means based on each of the "quasi-simultaneous" light curves 
by less than 0.041$^m$ and 0.075$^m$ in W1 and W2, respectively. However, the averages of the intensity means
computed separately for the two epoch intervals practically coincide with the corresponding intensity
means computed based on all available observations ignoring the Blazhko variations: the differences
do not exceed 0.015$^m$ and 0.010$^m$ in W1 and W2, respectively. In NGC5466 each star
has only two "outlying" measurements (about MJD 55203.080975), while the bulk of observations
(15 measurements) are concentrated within a $\sim$~1.1-day long interval (MJD 55380.630970 -- 55381.755488).
The inclusion/exclusion of the two "rogue" measurements has negligible effect on the final intensity means with
the differences not exceeding 0.009$^m$ and 0.050$^m$ in W1 and W2, respectively (the standard errors 
are greater than 0.037$^m$ and 0.074$^m$ in W1 and W2, respectively, for all the stars concerned). 
Given the smallness of the Blazhko-variation due effect on the inferred intensity means and the 
small fraction of Blazhko stars in our sample (15 out of 73-74 stars in M3, 3 out of 36 stars in M5, 
6 out of 28 stars in M15, 1 out of 58 stars in NGC3201, and 2 out of 9 stars in NGC5466 with no Blazhko stars 
in other clusters) hereafter we adopt the intensity mean W1- and W2-band magnitudes computed based on all available
WISE observations for all stars ignoring eventual light-curve variations.
Figure~\ref{blazhko} shows several examples of Blazhko star light curves in three clusters.

Table~\ref{clusters} lists the data for our calibrator  GCs including the number of RR Lyrae
found with adopted WISE W1- and W2-band light curves as well as the metallicity 
in the  scale of \citet{carretta} and the reddening  E(B-V), both adopted from the 
updated version of the  globular-cluster catalogue by \citet{harris} \citep{harris2}.

\begin{table}
 \centering
  \caption{The sample of calibrator GCs. Metallicities are in the CG scale. }
  \begin{tabular}{@{}llccc@{}}
  \hline
  Name     & [Fe/H] & E(B-V) & \multicolumn{2}{c}{Number of   RR Lyr} \\
           &        &        & (W1)                           & (W2) \\
 \hline
 M3           & -1.50 & 0.01  & 74 & 73 \\
 M4           & -1.16 & 0.35  & 31 & 31 \\
 M5           & -1.29 & 0.03  & 36 & 36 \\
 M15          & -2.37 & 0.10  & 28 & \\
 M53          & -2.10 & 0.02  & 22 & \\
 M55          & -1.94 & 0.08  &  6 & 6 \\
 M92          & -2.31 & 0.02  &  7 & 7 \\
 M107         & -1.02 & 0.33  &  9 & 9 \\
 NGC 3201     & -1.59 & 0.24  & 58 & 58 \\
 NGC 5053     & -2.27 & 0.01  &  8 & \\
 NGC 5466     & -1.98 & 0.00  &  9 & \\
 NGC 6362     & -0.99 & 0.09  & 17 & 17 \\
 NGC 6723     & -1.10 & 0.05  &  9 & \\
 NGC 6934     & -1.47 & 0.10  &  8 & \\
 $\omega$ Cen & -1.75$^*$ & 0.12 & 38 & 38 \\
 \hline
\end{tabular}
\hspace{2cm} $^*$ Because of the well known metallicity spread
 among RR Lyrae stars in this cluster (Sollima et al. 
2006 and reference therein), we took into account 
only the metal-poor ($[Fe/H]<-1.4$ )
$$ $$
\label{clusters}
\end{table}

\section{Calibration of the PML relation}
\label{calibr}

So far, three studies involved the estimation of the parameters  
of the period-metallicity-luminosity relation in the form of eq.~(\ref{PMLR}) for 
WISE mid-infrared photometric bands, all of them based on field stars. We summarise the corresponding
results in Table~\ref{Published_Studies}. Given eq.~(\ref{PMLR}), the apparent 
X-band magnitude of a particular star is equal to
\begin{equation}
\label{eq2}
<X>=\alpha_X~log P_{F}+\beta_X~[Fe/H]+\gamma_X+(m-M)_0 +A_{X},
\end{equation}
where $<X>$ is the intensity-mean X-band magnitude; $(m-M)_0$, the true distance modulus,
and $A_{X}$, the total extinction in the X band.

We now proceed to determine the three parameters (coefficients) $\alpha_{X}$, $\beta_{X}$,
and $\gamma_{X}$ of eq.~(\ref{PMLR}) for the two shortest-wavelengths WISE passbands
X=W1 and X=W2 from observational data.

\begin{table}
 \centering
  \caption{Published determinations of the parameters of the
RR Lyrae period-metallicity-luminosity relations in WISE photometric bands.}
  \begin{tabular}{@{}lcccl@{}}
  \hline
   Ref.  & Filter & $\alpha$ & $\beta$  & $\gamma$ \\
 \hline
 K11                               & W1 & -1.681 &       & +0.083\\
 K11                               & W2 & -1.715 &       & +0.092\\
 K11                               & W3 & -1.688 &       & +0.013\\
 D13                               & W1 & -2.33~ & 0.088 & -0.825\\
 M13                               & W1 & -2.44~ &       & -1.26~\\
 M13                               & W2 & -2.55~ &       & -1.29~\\
 M13                               & W3 & -2.58~ &       & -1.32~\\
 \hline
\end{tabular}
\label{Published_Studies}

References: K11: \citet{klein}; D13: \citet{dambis3}; M13: \citet{madore}. 
\end{table}

\subsection{The period slopes ($\alpha_{X}$)}
\label{period}

\begin{figure*}
 \includegraphics[width=17.cm]{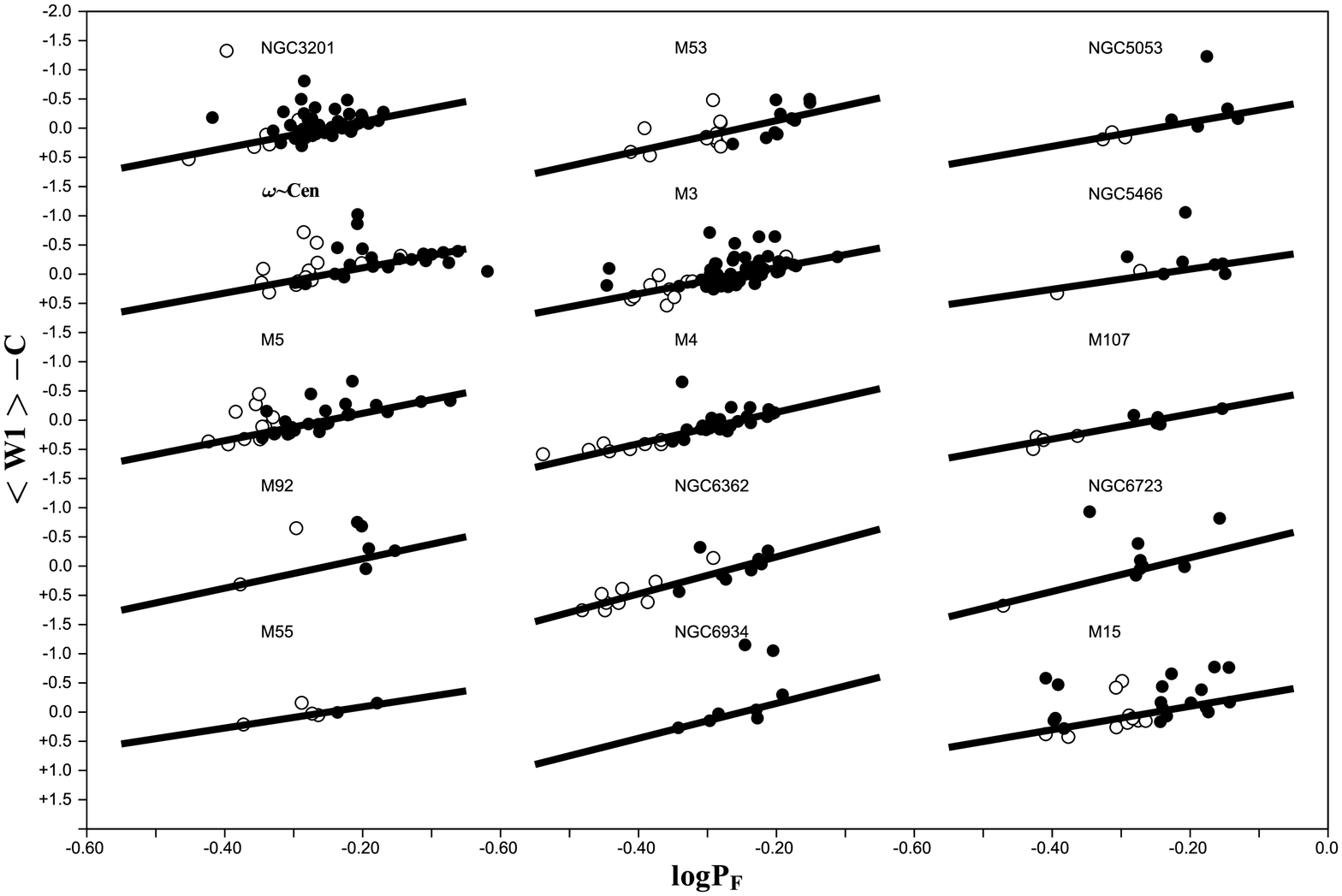}
\caption{The PL$_{W1}$ relation for the RR Lyrae in 15 calibrator globular clusters. 
The filled and open circles are the RRab and RRc type variables with fundamentalised periods, respectively. 
The W1 magnitudes are scaled to the same distance, extinction, and metallicity by subtracting the parameter 
$C_{W1}$ for each cluster.}
\label{all}
\end{figure*}

\begin{figure*}
 \includegraphics[width=17.cm]{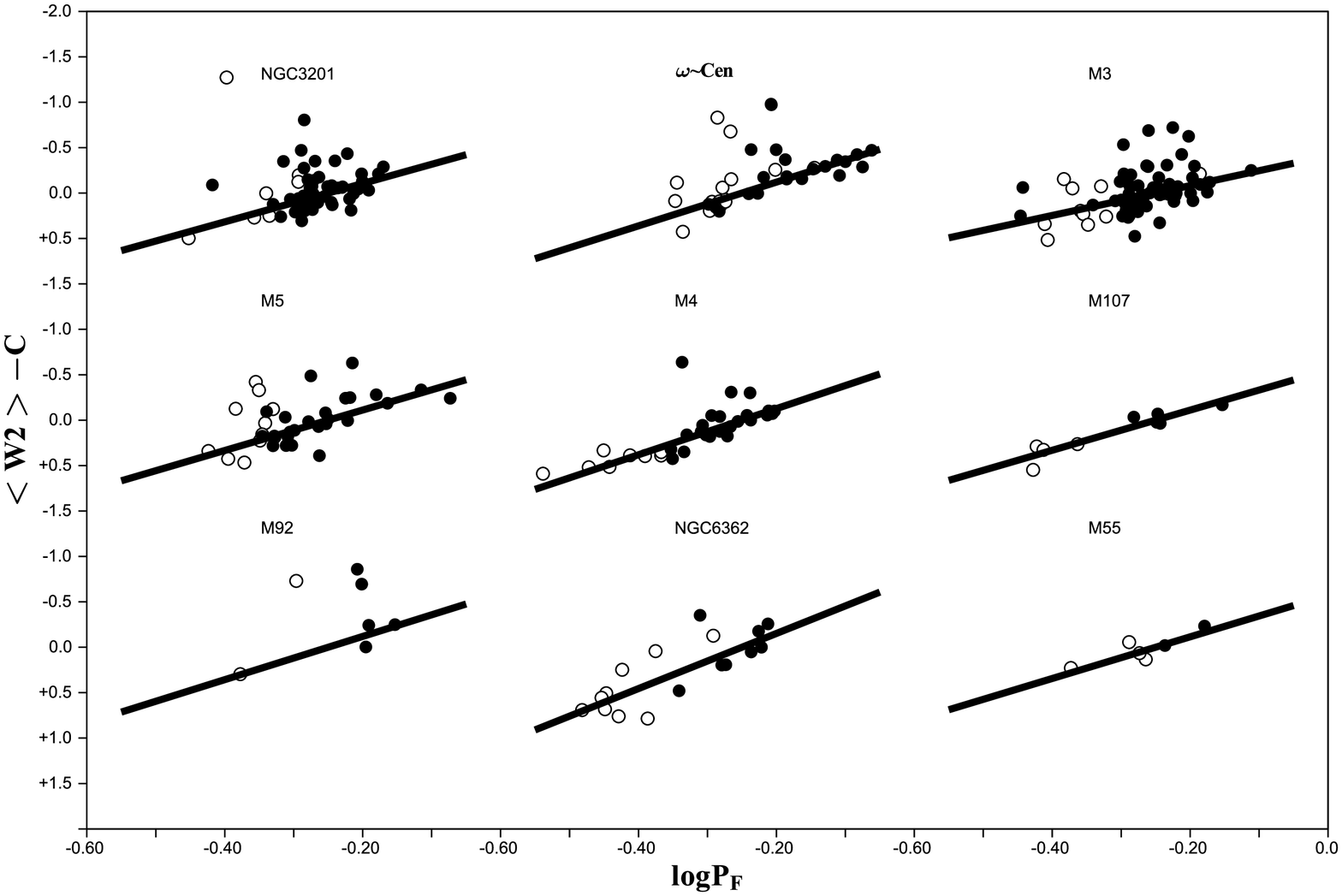}
\caption{The PL$_{W2}$ relation for the RR Lyrae in 9 calibrator globular clusters. 
The filled and open circles are the RRab and RRc type variables with fundamentalised periods, respectively. 
The W2 magnitudes are scaled to the same distance, extinction, and metallicity by subtracting the parameter 
$C_{W2}$ for each cluster.}
\label{all2}
\end{figure*}

All  stars in a particular cluster can be considered to be at the same distance
(which is much greater than the size of the cluster and hence the line-of-sight extent
of the system can be neglected) and (in most cases) to have the 
same metallicity and the same amount of interstellar extinction (anyway, intracluster 
extinction variations in all WISE photometric bands are at least about a factor of 17 smaller 
than the corresponding variations in the V-band extinction \citep{yuan} and therefore negligible).  
Equation (\ref{eq2}) for stars of a given cluster then acquires the form
\begin{equation}
\label{eq3}
<X>-\alpha_{X}~(log P_{F}+0.25)=C_X$$
\end{equation}
where $C_X = \beta_X~[Fe/H]+\gamma_X+(m-M)_0 +A_{X} -0.25 \alpha_{X}$ can be considered to be a constant.
Hereafter we add the +0.25 term to $log P_{F}$ in order to centre the solution at $log P_{F}$=-0.25, which
is close to the average value of this parameter, so as to make the $C_X$ constant more representative
of the cluster distance modulus and minimise the effect of differences in the inferred $\alpha_{X}$
values between different clusters.
We use the following heuristic procedure to  estimate the constant $C_X$ for some assumed slope $\alpha_{X}$. We compute
the left-hand side of eq. (\ref{eq3}) $c_{\alpha,i}$ = $<X>-\alpha_{X}~(log P_{F,i} + 0.25)$ for each star.
We then sort the  $C_{\alpha,i}$ values in the ascending order and 
seek the subset $\mu = \{j,j+1... j+N_1-1 \}$ containing $N_1 = N \times q$ values with $q$=0.68 (where N is the total 
number of RR Lyraes in the given cluster for which we determined the corresponding X-band
intensity mean magnitudes) having the smallest dispersion of computed $C_{\alpha}$ values,
$\sigma C_{\alpha, \mu}$ (we adopt $q$=0.8 for NGC5053 and M92 and $q$=1.0 for NGC6934). We then try
$\alpha$ values from -1.0 to -5.0 in increments of 0.01 to find the one yielding
the smallest $\sigma C_{\alpha, \mu}$. 
If the modal "core" distribution (i.e., the part of the distribution corresponding to stars
whose data points outline the purported linear log $P_{F}$-$<X>$ relation) of $C_i$ values were normal, 
our subset would roughly consist at least of
all stars with $C_j$ between the $<C> -\sigma C$ and $<C> +\sigma C$, where $<C>$ and
$\sigma C$ are the mean and dispersion of C values for the subset of stars defining
the linear log $P_{F}$-$<X>$ relation, respectively. The mean C value
averaged over the subset stars, $<C_{\mu}>$, should then be close to the
mean $<C>$, and the (truncated) dispersion $\sigma C_{\mu}$ should be roughly equal to
$\sigma C_{subset}$ =0.54$\sigma C$  and hence 3$\sigma C$ = 5.56$\sigma C_{subset}$.
We therefore determine the final estimate of C and $\alpha_{X}$ by least-squares 
solving the equation set
\begin{equation}
\label{eq4}
\alpha_{X}~(log P_{F}+0.25)+C = <X>,
\end{equation}
(it is just a rewritten form of eq.~(\ref{eq3})) for stars with $C_{\mu}$ values
in the interval  
$<C_{\mu}> - 5.56 \sigma C_{\mu} \leq C_{\mu} \leq <C_{\mu}> - 5.56 \sigma C_{\mu}$.
The resulting solutions (i.e., the $\alpha_{X}$ and $C_X$ values, their standard errors and the standard 
error of $<X>$, where X=W1 or W2) for all globular clusters, where such solutions could be reasonably derived,
are listed in Table~\ref{clustersolutions}. Like \citet{sollima}, we plot the scaled W1 and W2 magnitudes 
($W1-C_{W1}$ and $W2-C_{W2}$)for our calibrating clusters as a function of fundamentalised periods in 
Figs.~\ref{all} and \ref{all2}, respectively.

\begin{table*}
 \centering
  \caption{Parameters of the  $<W1>$=$\alpha_{W1}~(log P_{F}+0.25) + C_{W1}$ 
and $<W2>$=$\alpha_{W2}~(log P_{F}+0.25) + C_{W2}$ fits for the globular clusters of our sample. }
  \begin{tabular}{@{}lcccccc@{}}
  \hline
   Name       &    $\alpha_{W1}$    & $C_{W1}$         & $\sigma <W1>$& $\alpha_{W2}$ & $C_{W2}$ &  $\sigma <W2>$ \\
 \hline
 M3           &  -2.235 $\pm$ 0.256 & 14.410 $\pm$ 0.016 & 0.124    &   -1.642 $\pm$ 0.332 & 14.383 $\pm$ 0.021 & 0.161 \\
 M4           &  -2.694 $\pm$ 0.213 & 10.818 $\pm$ 0.020 & 0.082    &   -2.540 $\pm$ 0.248 & 10.817 $\pm$ 0.024 & 0.096 \\
 M5           &  -2.343 $\pm$ 0.236 & 13.770 $\pm$ 0.020 & 0.098    &   -2.225 $\pm$ 0.321 & 13.824 $\pm$ 0.027 & 0.134 \\
 M15          &  -2.013 $\pm$ 0.445 & 14.382 $\pm$ 0.036 & 0.156    &          &            &  \\
 M53          &  -2.588 $\pm$ 0.580 & 15.542 $\pm$ 0.044 & 0.193    &          &            &  \\
 M55          &  -1.817 $\pm$ 0.207 & 13.004 $\pm$ 0.013 & 0.025    &   -2.294 $\pm$ 0.553 & 12.997 $\pm$ 0.036 & 0.068 \\
 M92          &  -2.516 $\pm$ 0.969 & 13.766 $\pm$ 0.087 & 0.138    &   -2.379 $\pm$ 0.665 & 13.795 $\pm$ 0.060 & 0.095 \\
 M107         &  -2.158 $\pm$ 0.319 & 13.214 $\pm$ 0.035 & 0.083    &   -2.210 $\pm$ 0.314 & 13.196 $\pm$ 0.035 & 0.082 \\
 NGC 3201     &  -2.284 $\pm$ 0.306 & 12.791 $\pm$ 0.016 & 0.106    &   -2.112 $\pm$ 0.343 & 12.797 $\pm$ 0.018 & 0.118 \\
 NGC 5053     &  -2.071 $\pm$ 0.495 & 15.512 $\pm$ 0.038 & 0.089    &          &            &  \\
 NGC 5466     &  -1.729 $\pm$ 0.531 & 15.409 $\pm$ 0.045 & 0.103    &          &            &  \\
 NGC 6362     &  -3.167 $\pm$ 0.384 & 13.742 $\pm$ 0.051 & 0.138    &   -3.034 $\pm$ 0.544 & 13.803 $\pm$ 0.072 & 0.195 \\
 NGC 6723     &  -2.894 $\pm$ 0.583 & 14.012 $\pm$ 0.055 & 0.105    &          &            &  \\
 NGC 6934     &  -2.990 $\pm$ 0.898 & 15.459 $\pm$ 0.047 & 0.100    &          &            &  \\
 $\omega$ Cen &  -2.158 $\pm$ 0.197 & 13.115 $\pm$ 0.018 & 0.087    &   -2.409 $\pm$ 0.222 & 13.152 $\pm$ 0.020 & 0.098 \\
 \hline
\end{tabular}
\label{clustersolutions}
\end{table*}

\begin{figure}
 \includegraphics[width=8.7cm]{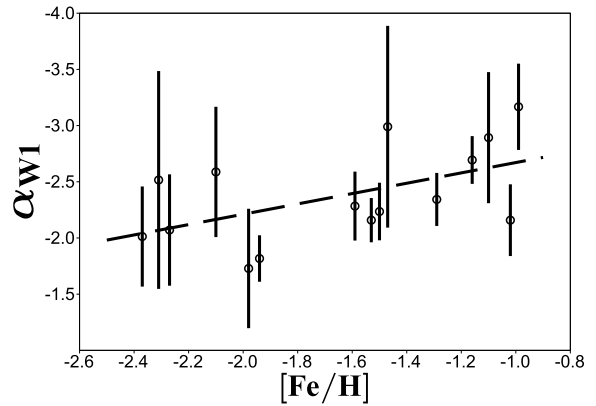}
\caption{The parameter $\alpha_{W1}=\delta M_{W1}/\delta logP$ for RR Lyrae stars 
as a function of the cluster metallicity. The dashed line shows the relation defined
by equation~(\ref{slopew1}). }
\label{pkmet}
\end{figure}

\begin{figure}
 \includegraphics[width=8.7cm]{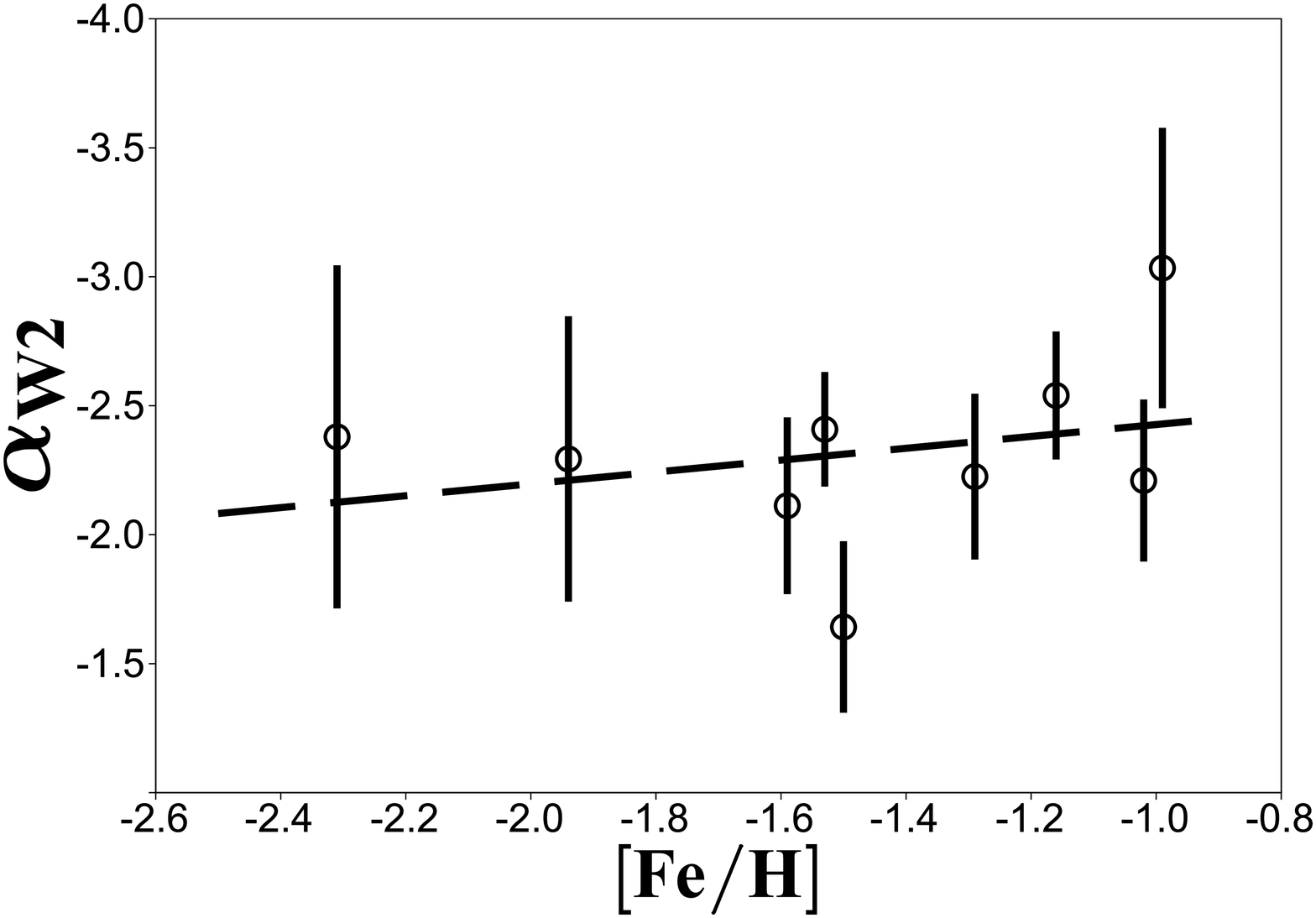}
\caption{The parameter $\alpha_{W2}=\delta M_{W2}/\delta logP$ for RR Lyrae stars 
as a function of the cluster metallicity. The dashed line shows the relation defined
by equation~(\ref{slopew2}).}
\label{pkmet2}
\end{figure}

\begin{figure}
 \includegraphics[width=8.7cm]{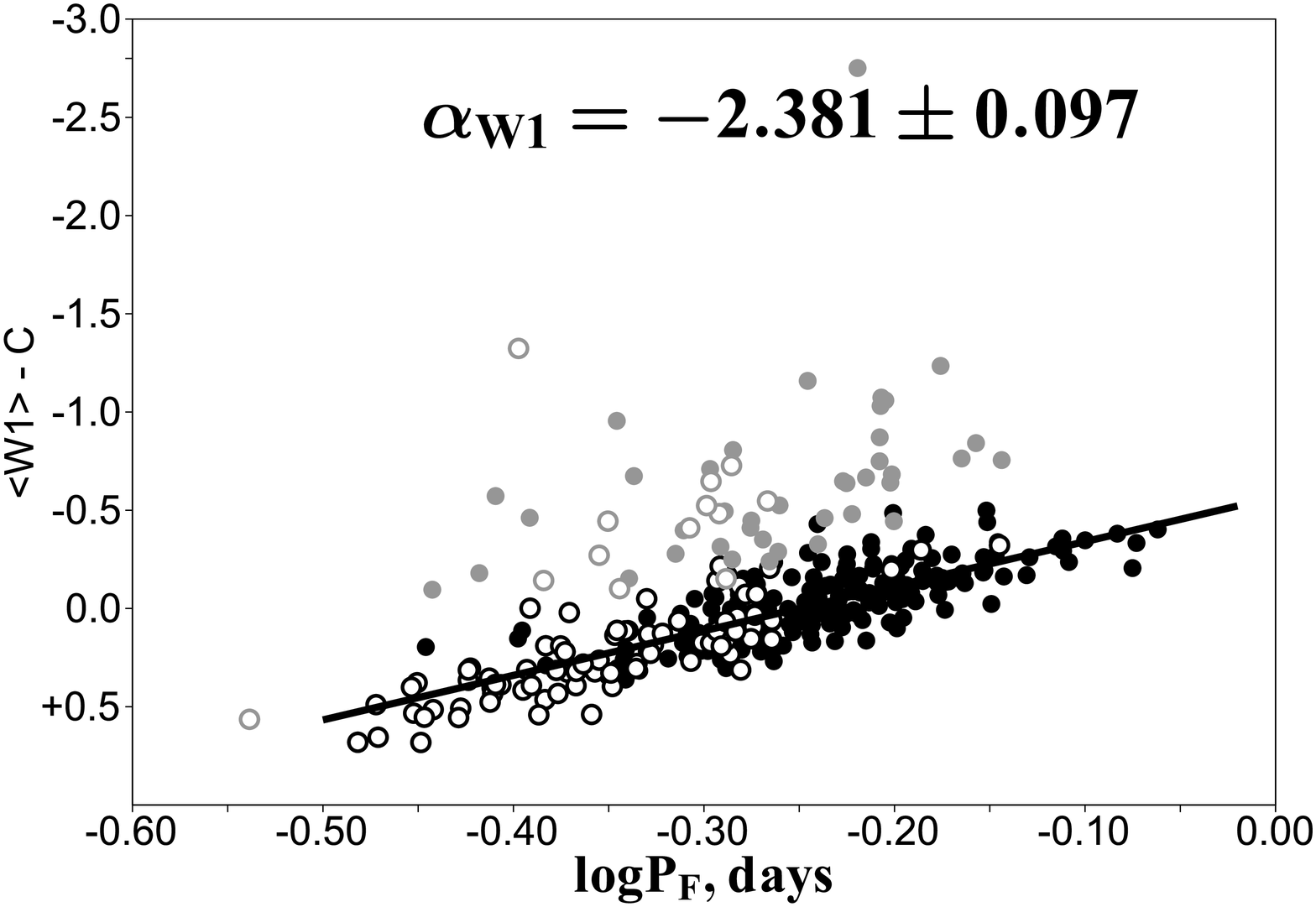}
\caption{The PL$_{W1}$ relation for the 360 RR Lyrae of our sample. The solid line 
shows the resulting fit. Filled circles are the 
RRab variables, open circles are the RRc variables whose periods have been 
fundamentalised. The gray symbols are the 3$\sigma$-rejected data points.}
\label{alpha}
\end{figure}

\begin{figure}
 \includegraphics[width=8.7cm]{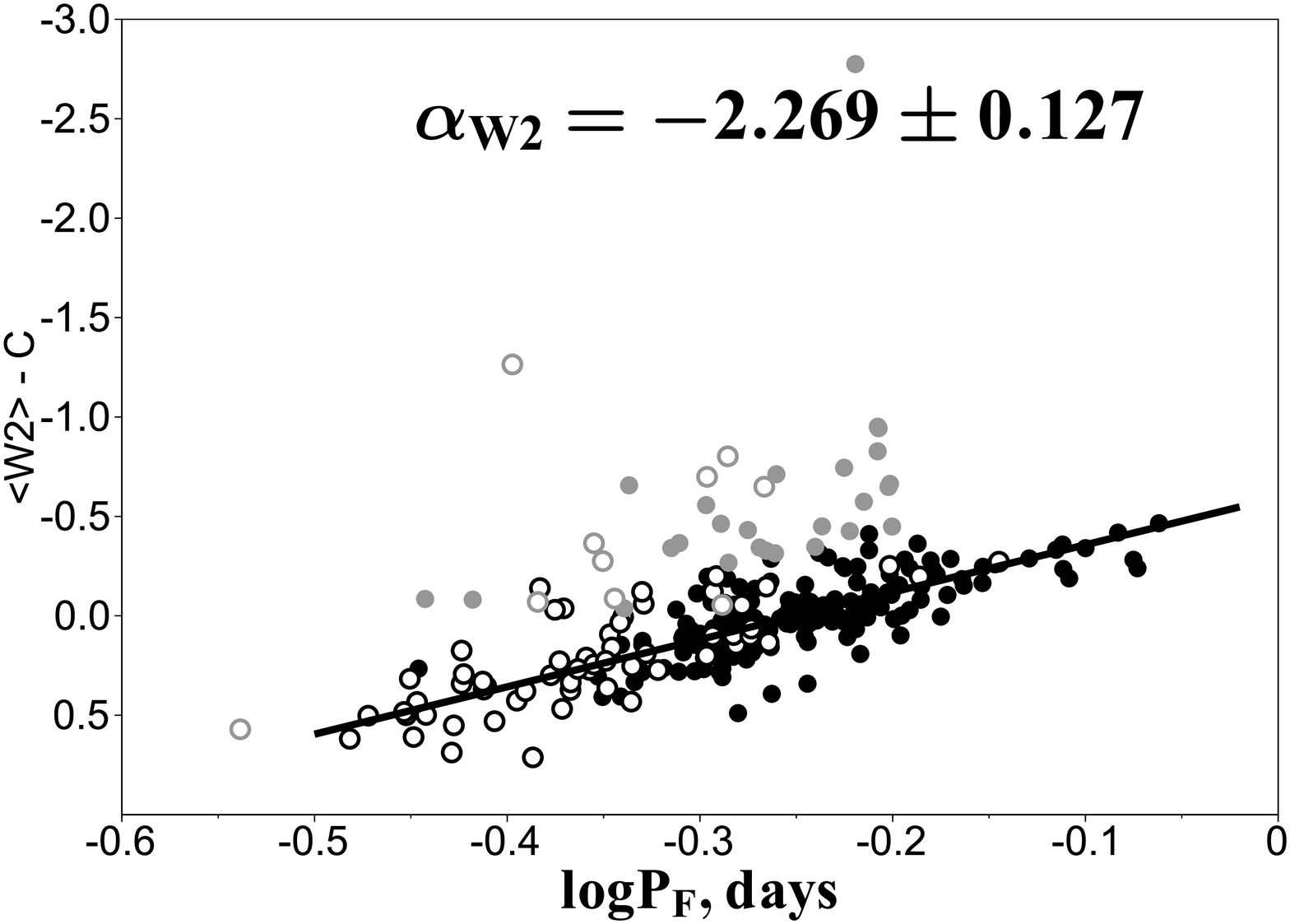}
\caption{The PL$_{W2}$ relation for the 275 RR Lyrae of our sample. The solid line 
shows the resulting fit. Filled circles are the 
RRab variables, open circles are the RRc variables whose periods have been 
fundamentalised. The gray symbols are the 3$\sigma$-rejected data points.}
\label{alpha2}
\end{figure}

Figures~\ref{pkmet} and \ref{pkmet2} show the individual cluster slopes $\alpha_{W1}$
and $\alpha_{W2}$ as a function of metallicity.
Linear least squares analysis yields the following results concerning
the possible metallicity dependence of the slopes $\alpha_{W1}$ and $\alpha_{W2}$: 
\begin{equation}
\label{slopew1}
\alpha_{W1} = -2.441 \pm 0.101 - (0.46 \pm 0.21) ([Fe/H]+1.5)
\end{equation}
and
\begin{equation}
\label{slopew2}
\alpha_{W2} = -2.311 \pm 0.127 - (0.23 \pm 0.31) ([Fe/H]+1.5).
\end{equation}
The slope $\alpha_{W2}$ appears to be independent of metallicity, whereas there seems to be
hint of a dependence  in the case of $\alpha_{W1}$. However, even in the latter case the slope differs 
from zero by less than  2.2$\sigma$ and we therefore derive the combined
solutions for both photometric bands (see Figs.~\ref{alpha} and \ref{alpha2}),
yielding the final slopes of $\alpha_{W1}$=-2.381~$\pm$~0.098 and $\alpha_{W2}$=-2.269~$\pm$~0.127.  
Table~\ref{singleslopesolutions} lists the resulting  $C_{W1}$ and $C_{W2}$ values obtained in terms of these
solutions (i.e., by forcing the same slope for all clusters).

\begin{table}
 \centering
  \caption{Parameters $C_{W1}$ and $C_{W2}$ of the combined  $<W1>= \alpha_{W1}~(log P_{F}+0.25) + C_{W1}$
and $<W2>=\alpha_{W2}~(log P_{F}+0.25) + C_{W2}$ single-slope fits for the globular clusters of our sample 
($\alpha_{W1}$=-2.381~$\pm$~0.098 and $\alpha_{W2}$=-2.269~$\pm$~0.127). } 
  \begin{tabular}{@{}lcc@{}}
  \hline
Name         &         $C_{W1}$      &      $C_{W2}$  \\
 \hline
M3           &    14.407 $\pm$ 0.015 &    14.369 $\pm$ 0.027 \\
M4           &    10.836 $\pm$ 0.024 &    10.834 $\pm$ 0.027 \\
M5           &    13.768 $\pm$ 0.022 &    13.822 $\pm$ 0.025 \\
M15          &    14.373 $\pm$ 0.027 &   \\                    
M53          &    15.544 $\pm$ 0.027 &   \\                    
M55          &    13.000 $\pm$ 0.054 &    12.997 $\pm$ 0.061 \\
M92          &    13.766 $\pm$ 0.061 &    13.792 $\pm$ 0.069 \\
M107         &    13.200 $\pm$ 0.041 &    13.192 $\pm$ 0.046 \\
NGC 3201     &    12.789 $\pm$ 0.017 &    12.794 $\pm$ 0.020 \\
NGC 5053     &    15.518 $\pm$ 0.046 &   \\                    
NGC 5466     &    15.425 $\pm$ 0.046 &   \\                    
NGC 6362     &    13.817 $\pm$ 0.032 &    13.876 $\pm$ 0.036 \\
NGC 6723     &    14.035 $\pm$ 0.050 &   \\                    
NGC 6934     &    15.466 $\pm$ 0.050 &   \\                    
$\omega$ Cen &    13.124 $\pm$ 0.022 &    13.146 $\pm$ 0.025 \\
 \hline
\end{tabular}
\label{singleslopesolutions}
\end{table}

\subsection{The metallicity slopes ($\beta_{X}$)}
\label{met}

We now follow the procedure employed by \citet{dambis3} to estimate the metallicity slopes $\beta_{W1}$
and $\beta_{W2}$  of the W1- and W2-band PML relations for RR Lyraes. The following analysis 
is to a large degree based on our previous paper \citep{dambis3}, where we use the metallicity scale
of \citet{ZW84}, and we therefore use metallicities on this scale throughout this subsection. We make the necessary 
transformation to the modern scale of \citet{carretta} in the next subsection.
We first compute the $(<V> - <W1>)_0$ and $(<V> - <W2>)_0$ intrinsic colour indices of 265
field RR Lyraes with $|b|~>25^o$ from Table~2 of \citet{dambis3} by dereddening the corresponding observed
$(<V> - <W1>)$ and $(<V> - <W2>)$ colours using the $A_V$ values from the above paper 
(computed using the 3D extinction map by \citet{drimmel}) and 
the reddening law by \citet{yuan} ($R_V =A_V/E_{B-V}$ = 3.1, $R_{W1} =A_{W1}/E_{B-V}$ = 0.18,
and $R_{W2} =A_{W2}/E_{B-V}$ = 0.16). We adopt
the $<V>$ and $<W1>$ intensity-mean magnitudes from the above paper and compute the $<W2>$ intensity-mean magnitudes
from WISE epoch photometry.
Like in our previous work, we proceed based on the following established facts.
First, the absolute V-band magnitude of RR Lyrae variables depends on metallicity [Fe/H] and,
for a given metallicity, is independent of period.  A consensus appears to have emerged 
concerning the slope of the [Fe/H]-$<M_V>$ relation for RR Lyraes. Thus Baade-Wesselink analyses yield
$\beta_V$~=~0.20~ \citep{cacciari92},  $\beta_V$~=~0.21~$\pm$~0.05 \citep{skillen}, 
and $\beta_V$~=~0.20~$\pm$~0.04 \citep{f2}, whereas \citet{gratton} and  \citet{federici} estimate the 
slope to be $\beta_V$~=0.214~$\pm$~0.047 and $\beta_V$~=0.25~$\pm$~0.02, respectively, 
based on observations of RR Lyraes in the LMC and horizontal-branch stars
in M31 globular clusters, respectively. Like in our previous study, we try to remain as "empiric" as possible and
therefore  we adopt the simple (unweighted) average of the latter two estimates
\begin{equation}
<M_V> = \gamma_{V} + 0.232 (\pm~0.020) \cdot [Fe/H]_{ZW},
\label{MVF}
\end{equation}
because they 
are based on the sole geometric assumption that the stars involved in both cases are practically at the same distance from 
us. 
Second, given the $\alpha_{W1}$=-2.381~$\pm$~0.097 and $\alpha_{W2}$=-2.269~$\pm$~0.127 slopes derived above,
the W1- and W2-band PML relations for RR Lyraes have the form:
\begin{equation}
<M_{W1}> = \gamma_{W1} + \beta_{W1} \cdot [Fe/H]_{ZW} -2.381 \cdot log P_F
\label{MW1}
\end{equation}
and
\begin{equation}
<M_{W2}> = \gamma_{W2} + \beta_{W2} \cdot [Fe/H]_{ZW} -2.269 \cdot log P_F,
\label{MW2}
\end{equation}
respectively. 
We then subtract equations~(\ref{MW1}) and (\ref{MW2}) from equation~(\ref{MVF}) to obtain:
$$
(<V>-<W1>)_0 = <M_V> - <M_{W1}> = 
$$
\begin{equation}
= (\gamma_{V} - \gamma_{W1}) + (0.232 - \beta_{W1}) \cdot [Fe/H]_{ZW} + 2.381 \cdot log P_F
\label{VW1}
\end{equation}
and 
$$
(<V>-<W2>)_0 = <M_V> - <M_{W2}> = 
$$
\begin{equation}
= (\gamma_{V} - \gamma_{W2}) + (0.232 - \beta_{W2}) \cdot [Fe/H]_{ZW} + 2.269 \cdot log P_F,
\label{VW2}
\end{equation}
respectively. We finally subtract the terms $2.381 \cdot log P_F$ and $2.269 \cdot log P_F$ from both sides of
equations~(\ref{VW1}) and (\ref{VW2}) to obtain:
$$
(<V>-<W1>)_0 - 2.381 \cdot log P_F = 
$$
\begin{equation}
= (\gamma_{V} - \gamma_{W1}) + (0.232 - \beta_{W1}) \cdot [Fe/H]_{ZW} 
\label{VW1a}
\end{equation}
and 
$$
(<V>-<W2>)_0 - 2.269 \cdot log P_F =
$$
\begin{equation}
= (\gamma_{V} - \gamma_{W2}) + (0.232 - \beta_{W2}) \cdot [Fe/H]_{ZW} 
\label{VW2a}
\end{equation}
Our calibrating stars now are 265  field RR Lyraes from Table~2 from 
\citet{dambis3} located at Galactic latitudes $|b| \ge +25^o$ and
with known $V$-, $W1$-, and $W2$-band intensity mean magnitudes. 
We finally solve equations~(\ref{VW1a}) and (\ref{VW2a}) for parameters
$(\gamma_{V} - \gamma_{W1})$, $(0.232 - \beta_{W1})$ and $(\gamma_{V} - \gamma_{W2})$, 
$(0.232 - \beta_{W2})$, respectively, to find:
$$
(<V>-<W1>)_0 =
$$
\begin{equation}
= 1.908(\pm 0.019) + 0.126 (\pm 0.012) \cdot [Fe/H]_{ZW} +2.381 \cdot log P_F
\label{VW1b}
\end{equation}
with a scatter of 0.087, and 
$$
(<V>-<W2>)_0 =
$$
\begin{equation}
1.853(\pm 0.018) + 0.113 (\pm 0.012) \cdot [Fe/H]_{ZW} +2.269 \cdot log P_F
\label{VW2b}
\end{equation}
with a scatter of 0.083, implying $\beta_{W1}$~=~0.106~$\pm$~0.023 and 
$\beta_{W2}$~=~0.119~$\pm$~0.023, respectively.

\subsection{The zero points ($\gamma_{X}$)}
\label{zeropoint}

 Given our recent statistical-parallax calibration of the [Fe/H]-$<M_V>$ relation
\citep{dambis3}:
\begin{equation}
<M_V> = +1.094 (\pm 0.091) + 0.232 (\pm 0.020) \cdot [Fe/H]_{ZW},
\end{equation}
we immediately obtain the following RR Lyrae PML relations in the W1 and W2 bands:
$$
<M_{W1}> = -0.814 (\pm 0.093)
$$
\begin{equation}
 - 2.381(\pm 0.097) \cdot log P_F+ 0.106 (\pm 0.023) \cdot [Fe/H]_{ZW}
\label{W1statZW}
\end{equation}
and
$$
<M_{W2}> = -0.759 (\pm 0.093)
$$
\begin{equation}
 - 2.269(\pm 0.127) \cdot log P_F+ 0.119 (\pm 0.023) \cdot [Fe/H]_{ZW}.
\label{W2statZW}
\end{equation}
A transformation to the modern metallicity scale via equation 
\begin{equation}
[Fe/H]_{Carretta} = 1.105 [Fe/H]_{ZW} +0.160
\end{equation}
\citep{carretta} yields:
$$
<M_{W1}> = -0.829 (\pm 0.093) 
$$
\begin{equation}
- 2.381(\pm 0.097) \cdot log P_F+ 0.096 (\pm 0.021) \cdot [Fe/H]_{Carretta}
\label{W1stat}
\end{equation}
and
$$
<M_{W2}> = -0.776 (\pm 0.093)
$$
\begin{equation}
 - 2.269(\pm 0.127) \cdot log P_F+ 0.108 (\pm 0.021) \cdot [Fe/H]_{Carretta}.
\label{W2stat}
\end{equation}

We perform another calibration of the zero points $\gamma_{W1}$ and $\gamma_{W2}$
based on intensity-mean W1- and W2-band magnitudes and 
HST FSG trigonometric parallaxes of four RR Lyraes adopted from \citet{madore} and 
\citet{b3}, respectively, $\gamma_{W1, HST} = -1.135 \pm 0.077$ and $\gamma_{W2, HST} = -1.088 \pm 0.077$ 
for the metallicity scale  of \citet{ZW84} and $\gamma_{W1, HST} = -1.150 \pm 0.077$ and 
$\gamma_{W2, HST} = -1.105 \pm 0.077$ for the metallicity scale of \citet{carretta}. Hence 
the HST trigonometric-parallax based calibrations are:
$$
<M_{W1, HST}> = -1.135 (\pm 0.077) 
$$
\begin{equation}
- 2.381(\pm 0.097) \cdot log P_F+ 0.106 (\pm 0.023) \cdot [Fe/H]_{ZW}
\end{equation}
and
$$
<M_{W2, HST}> = -1.088 (\pm 0.077)
$$
\begin{equation}
 - 2.269(\pm 0.127) \cdot log P_F+ 0.117 (\pm 0.023) \cdot [Fe/H]_{ZW}.
\end{equation}
Or, for the metallicity scale of \citet{carretta}:
$$
<M_{W1, HST}> = -1.150 (\pm 0.077)
$$
\begin{equation}
\label{W1HST}
 - 2.381(\pm 0.097) \cdot log P_F+ 0.096 (\pm 0.021) \cdot [Fe/H]_{Carretta}
\end{equation}
and
$$
<M_{W2, HST}> = -1.105 (\pm 0.077) 
$$
\begin{equation}
\label{W2HST}
- 2.269(\pm 0.127) \cdot log P_F+ 0.108 (\pm 0.021) \cdot [Fe/H]_{Carretta}.
\end{equation}
The HST based distance scales can be seen to be longer than the statistical-parallax
based ones by 0.321 and 0.329 in terms of distance moduli for
the $PML_{W1}$ and $PML_{W2}$ relations, respectively. The discrepancy between the 
HST  and statistical-parallax distance scales appears to be important,
amounting to $\sim$~2.7~$\sigma$ in both cases.

Interestingly, a recent statistical-parallax calibration of the intensity-mean
$V$-band absolute magnitude ($<M_V>$) of RR Lyrae c-type variables by \citet{k12} 
yields $<M_V>$~=~0.59~$\pm$~0.10 at [Fe/H]=-1.59, which is $\sim$~0.14
brighter than our statistical-parallax based estimate \citep{dambis3} and therefore implies
the $\gamma_{W1}$ and $\gamma_{W2}$ estimates lying almost halfway between
those inferred from our statistical-parallax solution and from HST FSG trigonometric parallaxes.
The corresponding $\gamma_{W1}$ and $\gamma_{W2}$ zero points prove to be $\sim~\sigma$ brighter than
those implied by our calibration and $\sim~1.4\sigma$ fainter than those implied by HST parallaxes and, perhaps,
could reconcile the two. Note, however, that, unlike the study of \citet{k12}, which concerns RRc-type variables 
exclusively distributed mostly in the southern part of the sky and is based on the data for 242 stars, our statistical-parallax
analysis involves 387 stars representing a natural mix of RRab and RRc type variables distributed pole-to-pole throughout the entire sky.

\begin{table*}
 \centering
  \caption{Estimated distances  to the calibrator clusters.}
  \begin{tabular}{@{}lccccc@{}}
  \hline
   Name       & $DM_0$ ($PML_{W1}$) & $DM_0$ ($PML_{W2}$)  & $DM_0$ ($PML_{W1}$) & $DM_0$ ($PML_{W2}$)  & $DM_0$ ($PML_{K}$) \\
              & \multicolumn{2}{c}{statistical~parallax}  & \multicolumn{2}{c}{HST trigonometric} &  \citep{sollima} \\
              & \multicolumn{2}{c}{zero~point}  & \multicolumn{2}{c}{parallax zero point}  &  \\
 \hline 
 M3           & 14.78 $\pm$ 0.02  &   14.74 $\pm$ 0.03 & 15.10 $\pm$ 0.02 &  15.07 $\pm$ 0.02  & 15.07 \\
 M4           & 11.12 $\pm$ 0.02  &   11.11 $\pm$ 0.03 & 11.44 $\pm$ 0.02 &  11.44 $\pm$ 0.03  & 11.39 \\
 M5           & 14.12 $\pm$ 0.02  &   14.17 $\pm$ 0.03 & 14.44 $\pm$ 0.02 &  14.50 $\pm$ 0.03  & 14.35 \\
 M15          & 14.82 $\pm$ 0.03  &                    & 15.14 $\pm$ 0.03 &                    & 15.13 \\
 M53          & 15.98 $\pm$ 0.03  &                    & 16.30 $\pm$ 0.03 &                    &       \\
 M55          & 13.41 $\pm$ 0.05  &   13.40 $\pm$ 0.06 & 13.73 $\pm$ 0.05 &  13.73 $\pm$ 0.06  & 13.62 \\
 M92          & 14.22 $\pm$ 0.06  &   14.25 $\pm$ 0.07 & 14.55 $\pm$ 0.06 &  14.58 $\pm$ 0.07  & 14.65 \\
 M107         & 13.47 $\pm$ 0.04  &   13.45 $\pm$ 0.05 & 13.79 $\pm$ 0.04 &  13.78 $\pm$ 0.05  & 13.76 \\
 NGC 3201     & 13.13 $\pm$ 0.02  &   13.14 $\pm$ 0.02 & 13.45 $\pm$ 0.02 &  13.47 $\pm$ 0.02  & 13.40 \\
 NGC 5053     & 15.97 $\pm$ 0.04  &                    & 16.29 $\pm$ 0.04 &                    &       \\
 NGC 5466     & 15.85 $\pm$ 0.04  &                    & 16.17 $\pm$ 0.04 &                    &       \\
 NGC 6362     & 14.13 $\pm$ 0.03  &   14.18 $\pm$ 0.04 & 14.45 $\pm$ 0.03 &  14.51 $\pm$ 0.04  & 14.44 \\
 NGC 6723     & 14.37 $\pm$ 0.05  &                    & 14.69 $\pm$ 0.05 &                    &       \\
 NGC 6934     & 15.82 $\pm$ 0.05  &                    & 16.14 $\pm$ 0.05 &                    &       \\
 $\omega$ Cen & 13.50 $\pm$ 0.02  &   13.52 $\pm$ 0.03 & 13.82 $\pm$ 0.02 &  13.85 $\pm$ 0.03  & 13.72 \\
 \hline
\end{tabular}
\label{dist}
\end{table*}

\section{The Distances to the calibrator clusters}

We estimate the distance moduli of the calibrating clusters using 
the above PML relations with the zero points based both on 
the statistical-parallax solution (equations~(\ref{W1stat}) and (\ref{W2stat}))
and on HST trigonometric parallaxes (equations~(\ref{W1HST}) and (\ref{W2HST})).
The results are listed in Table~\ref{dist}, where the last column
gives the distance moduli estimated by \citet{sollima} based on the $PML_{K}$ relation. 
We find our cluster distance moduli based on the $PML_{W1}$ and $PML_{W2}$ relations
to be highly consistent with each other with the $<DM_0 (PML_{W1}) - DM_0 (PML_{W2})>$=-0.01~$\pm$~0.03
and  $<DM_0 (PML_{W1}) - DM_0 (PML_{W2})>$=-0.02~$\pm$~0.03 if computed with the
zero points tied to our statistical-parallax solution and HST FSG trigonometric parallaxes, respectively. 
Furthermore, our cluster distance estimates computed with HST based zero points agree well
with those found by \citet{sollima} using their derived the $PML_{K}$ relation with the
average  distance-modulus differences (this paper minus \citet{sollima}) of +0.04 and +0.06 and a scatter of 0.06.

\section{Conclusions}

Our analysis of WISE W1- and W2-band epoch photometry for 372 RR Lyrae type variables
in 15  Galactic globular clusters combined with V-band and WISE W1- and W2-band photometry
of 265 field RR Lyraes  at Galactic latitudes $|b|~>25^o$ allowed us to derive 
the period-metallicity-luminosity relations in the W1 and W2 bands. We derive two 
sets of appreciably discrepant zero points with one based on our recent statistical-parallax analysis \citep{dambis3}
and another one tied to the trigonometric parallaxes of four RR Lyraes measured with the
HST FGS \citep{b3}. The statistical-parallax based calibration yields
zero points that are 0.32$^m$ (W1) and 0.33$^m$ (W2) shorter than those calibrated with HST FGS 
parallaxes. 
The $\sim~0.3^m$ difference in the zero points given by two geometric methods 
is by no means trivial, but this is  long-standing issue, which still remains unresolved.
A more detailed discussion can be found in Section~6.1 of our previous paper~\citep{dambis3}. Let us hope
that GAIA will soon resolve the controversy.

We use our calibrations to estimate the distance moduli to 15 calibrator globular clusters
of which nine have distance determined using both $PML_{W1}$ and $PML_{W2}$ relations.
Our distances based on HST zero points agree well with the results of
\citet{sollima} with +0.04 and +0.06 distance-modulus differences both for $PML_{W1}$ and $PML_{W2}$ and the
scatter of 0.06 for the W1- and W2-based estimates, respectively.

\section*{acknowledgements}
We thank the anonymous reviewer for the valuable comments, which greatly improved the
final version of the paper. This publication makes use of data
products from  the Wide-field Infrared Survey Explorer, which is a 
joint project of the University of California, Los Angeles, 
and the Jet Propulsion Laboratory/California Institute of 
Technology, funded by the National Aeronautics and Space Administration. 
This research has made use of NASA's Astrophysics Data System.
This work is supported by the Russian Foundation for Basic Research
(projects nos.~13-02-00203-a and 11-02-00608-a).

\label{lastpage}

\end{document}